\documentclass{elsart}

\usepackage{natbib}

\usepackage{graphicx}

\usepackage{amssymb,amsmath}

\begin{document}

\begin{frontmatter}



\title{Parabolic dunes in north-eastern Brazil}


\author[1]{O.~Dur\'an \corauthref{cor1}}
\ead{o.duran@tudelft.nl}
\author[2]{M.V.N.~Silva}
\author[3]{L.J.C.~Bezerra}
\author[4,5]{H.J.~Herrmann}
\author[3]{L.P.~Maia}

\corauth[cor1]{Corresponding author. Permanent address: Nanostructured Materials (NSM), DelftChemTech, Julianalaan 136, 2628 BL Delft, The Netherlands. Tel.: +31 (0)15 278 5501.}

\address[1]{Institute for Computational Physics, Universit\"at Stuttgart, Pfaffenwaldring 27, D-70569 Stuttgart, Germany}
\address[2]{Universidade Federal do Pernambuco, Dept. Geos., Moraes R\^ego 1235, CEP 50670-901, Recife, Brazil}
\address[3]{Universidade Federal Ceara, Dept. Geol., Campus Pici Bloco 912, BR-60455970 Fortaleza, Cear\'a, Brazil}
\address[4]{Computational Physics, IfB, HIF E12, ETH H\"onggerberg, CH-8093 Z\"urich, Switzerland}
\address[5]{Departamento de F\'{\i}sica, Universidade Federal do Cear\'a, 60451-970 Fortaleza, Cear\'a, Brazil}

\begin{abstract}
In this work we present measurements of vegetation cover over parabolic dunes with different degree of activation along the north-eastern Brazilian coast. We are able to extend the local values of the vegetation cover density to the whole dune by correlating measurements with the gray-scale levels of a high resolution satellite image of the dune field. The empirical vegetation distribution is finally used to validate the results of a recent continuous model of dune motion coupling sand erosion and vegetation growth.
\end{abstract}


\end{frontmatter}

\section{Introduction}
\label{introduction}

The problems posed by dune mobility have been solved in practice using different techniques. Small dunes can be mechanically flattened so that sand moves as individual grains rather than as a single body. However, such methods are too expensive for large dunes. These can be immobilized covering them with oil or by the erection of fences. These solutions have the drawback of not providing a long term protection since the sand remains exposed. To overcome this shortcoming, a suitable solution is to vegetate the sand covered areas in order to prevent sediment transport and erosion \citep{Pye90}. This is particularly important for coastal management where a strong sand transport coexists with favorable conditions for vegetation growth. 

The stabilization of mobile sand using vegetation is an ancient technique. This method has been used with excellent results on coastal dunes in Algeria, Tunisia, North America, United Kingdom, Western Europe, South Africa, Israel among others \citep{Pye90}. Vegetation tries to stabilize sand dunes \citep{Hack41,Pye82,Anthonsen96,Muckersie95,Hesp96}, preventing sand motion \citep{Lancaster98} and stimulating soil recovery \citep{Tsoar90,Danin91}. 

Recently, we developed a mathematical description of the competition between vegetation growth and sand transport \citep{Duran06b}. This model was capable to reproduce the transformation of active barchan dunes into parabolic dunes under the action of vegetation growth as can be found in real conditions. Since numerical simulations are orders of magnitude faster than the real evolution, we are able to study the entire inactivation process and to forecast thousands of years of real evolution. 

Parabolic dunes are vegetated dunes that, when active, migrate along the prevailing wind direction. They arise under uni-directional wind and in places partially covered by plants and have a typical $U$ shape with the `nose' pointing downwind and the two arms pointing upwind, contrary to barchan dunes where the horns point downwind (Fig.~1). 
Vegetation covers most of the arms of parabolic dunes and a fraction of their nose depending on the activation degree of the dune, i.e. how fast the dune moves. An active parabolic dune has a sandy nose (Fig.~1, left) while an inactive one is almost totally covered by plants (Fig.~1, right). Plants are typically placed along the lee size of the dune, which is protected from wind erosion. On the contrary, the interior side exposed to the wind is devoid of vegetation. There, erosion is strong enough to prevent vegetation growth.

\begin{figure}[htb]
\centering
\fboxrule=0.5mm
\includegraphics[width=1.0\textwidth]{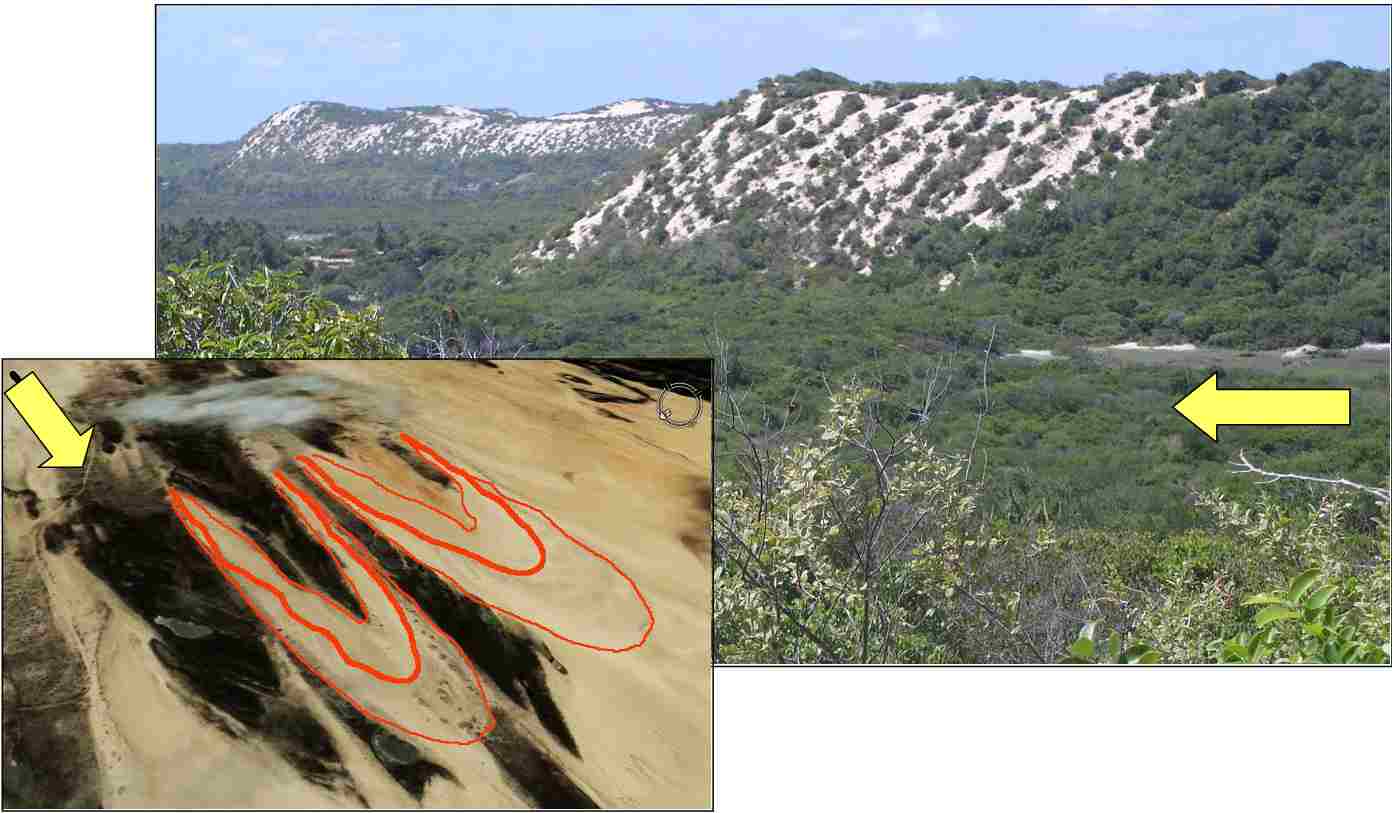}
\caption{(Color online) {\bf Left:} Satellite image of active parabolic dunes (delineated in red) in Pecem along the Brazilian coast. These dunes are partly covered by grass (in dark). {\bf Right:} Noses of marginally active parabolic dunes in Iquape along the Brazilian coast. These dunes are hundreds years old and can reach heights up to 50 m. The vegetation that covers them consist mainly of trees and shrubs. Arrows represent the wind direction.
}
\end{figure}

The migrating velocity of parabolic dunes is several times smaller than that of barchan dunes, and in general, they have an intermediate shape between fully active crescent dunes, like barchans, and completely inactive parabolic dunes. The activation degree of the parabolic dune is characterized by the vegetation cover pattern over it, which gives information about the areas of sand erosion and deposition pattern responsible for the motion of the dune.

In this work we study the degree of activation of parabolic dunes in north-eastern Brazil by direct measurements of the vegetation cover on them. We further present a method to extend the local information about vegetation cover to the whole dune by comparing the measured vegetation density cover with the gray scale level of high resolution satellite images. The empirical vegetation cover is finally quantitatively compared to the numerical solution of an established model for sand transport coupled with vegetation growth.

\section{Regional setting}

Along all the coast of the province of Cear\'a in the north-east of Brazil (Fig.~2), sand dunes are totally or partially stabilized by vegetation. On one hand, the humid climate of the region, with intense precipitation during the rain season from February to July (Fig.~3a), amplifies the role of vegetation as an active agent in the sandy landscape evolution. 

\begin{figure}[htb!]
\centering
\includegraphics[width=0.7\textwidth]{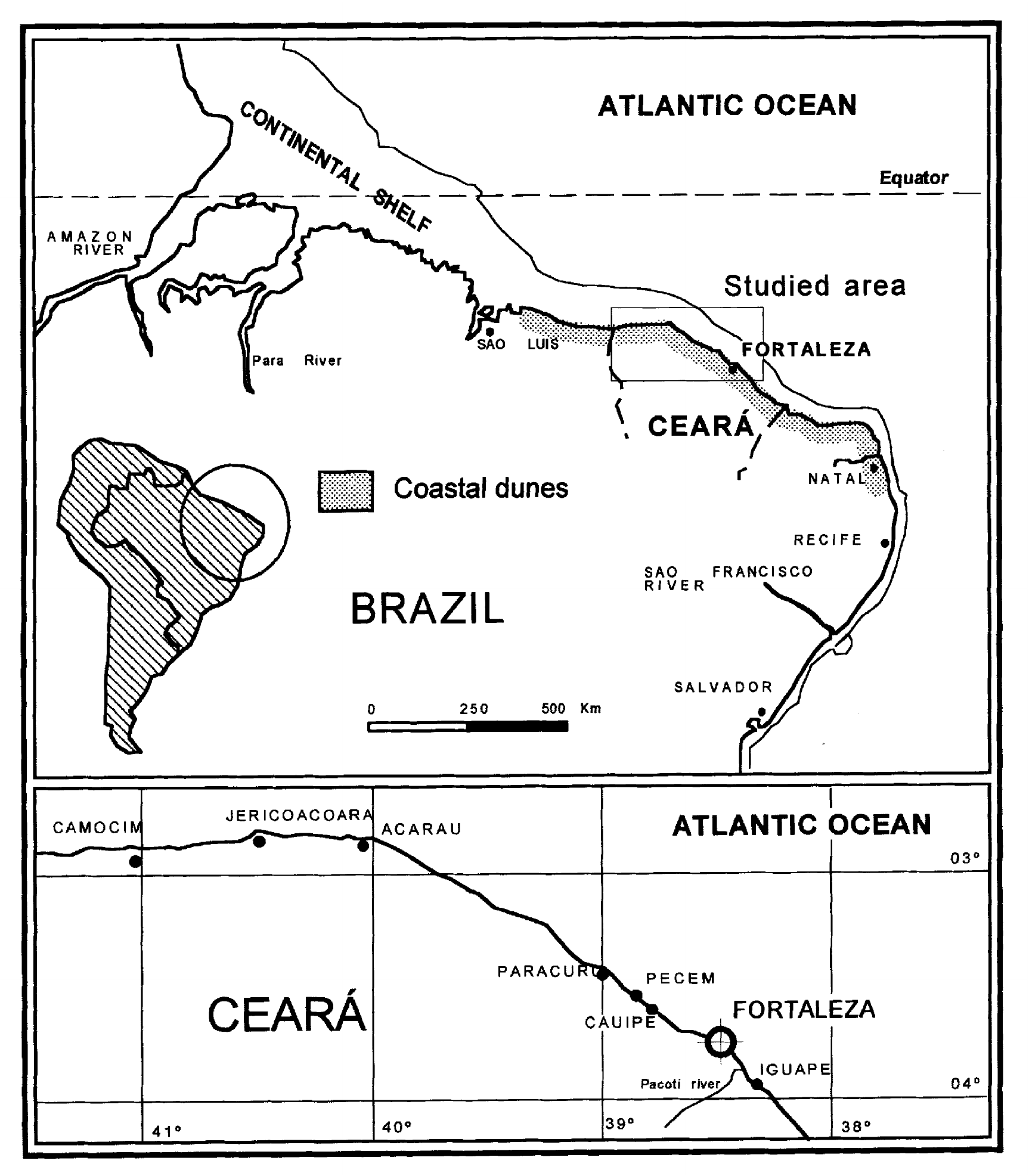}
\caption{Map of the north-eastern Brazil province of Cear\'a illustrating the studied region. We measured parabolic dunes along the coast near the city of Fortaleza, between the town of Iguape and Paracuru.}
\end{figure}

\begin{figure}[htb]
\centering
\includegraphics[width=0.8\textwidth]{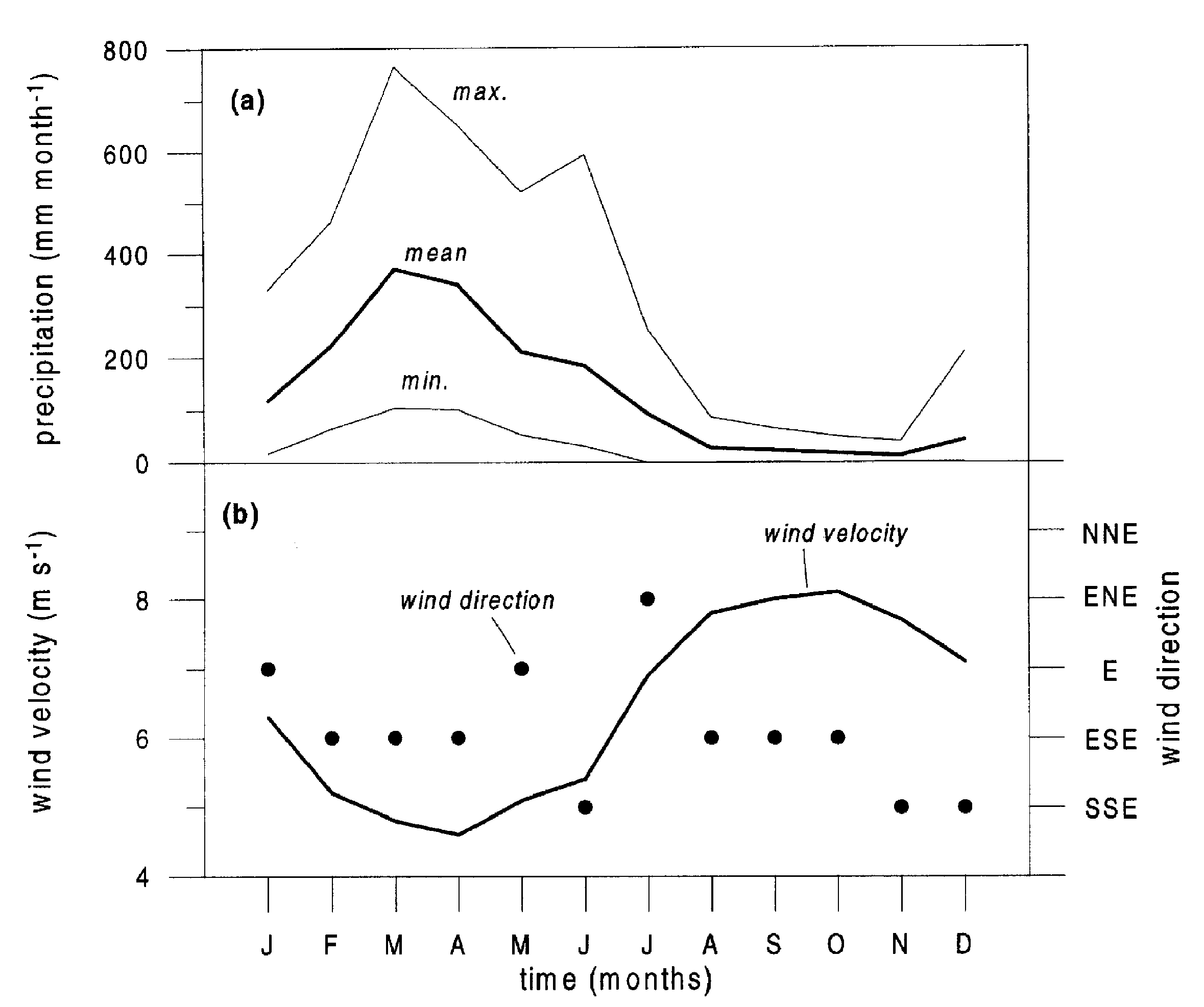}
\caption{{\bf (a)} Mean, maxima and minima monthly precipitation levels in the Fortaleza region. As most equatorial-sub-tropical climates, the year is clearly divided into a rain season, which here runs from February to July, and a dry season. {\bf (b)} Wind direction and velocity averaged over one month. The coastal wind is highly uni-directional, blowing from ESE most of the year. Notice that the oscillation of precipitation and wind velocity values have opposite phase.}
\end{figure}

On the other hand, the ubiquity of beaches plenty of sediments combined with a strong and highly uni-directional ESE coastal wind (Fig.~3b), create favorable conditions for the evolution of crescent sand dunes. Furthermore, since the wind is stronger during the dry season (Fig.~3), both processes, the aeolian sediment transport and the biomass production, lead to a competing effect that completely reshapes the coastal landscape by the development of parabolic dunes through the inversion of barchans and their further deactivation by the vegetation growth.

In order to obtain information about the distribution of vegetation over coastal dunes, we went to Fortaleza during the rain season to measure the shape of some parabolic dunes and the vegetation that covers them (Fig.~4). These dunes are located in Iguape, on the east of Fortaleza, and Pecem, Taiba and Paracuru, on the west (Fig.2). The geographical coordinates of all points are recorded with GPS and inserted in the digital dune map.

By using satellite Landsat images and a topographic map we were able to select parabolic dunes with different degrees of inactivation and vegetation cover density. Figures 4a, b, c, and d, show, in order of activation, the four measured parabolic dunes on the west coast of Fortaleza, while Figs.~4e, f, and g, show the other three dunes from the Iguape region, on the east coast. In general, the most active ones were located in Taiba and Iguape (shown in Figs.~4d and g, respectively), while those in Paracuru (Figs.~4a and b) were among the most inactive ones. 

\begin{figure}[htb!]
\centering
\fboxrule=0.5mm
\framebox{\includegraphics[width=1.0\textwidth]{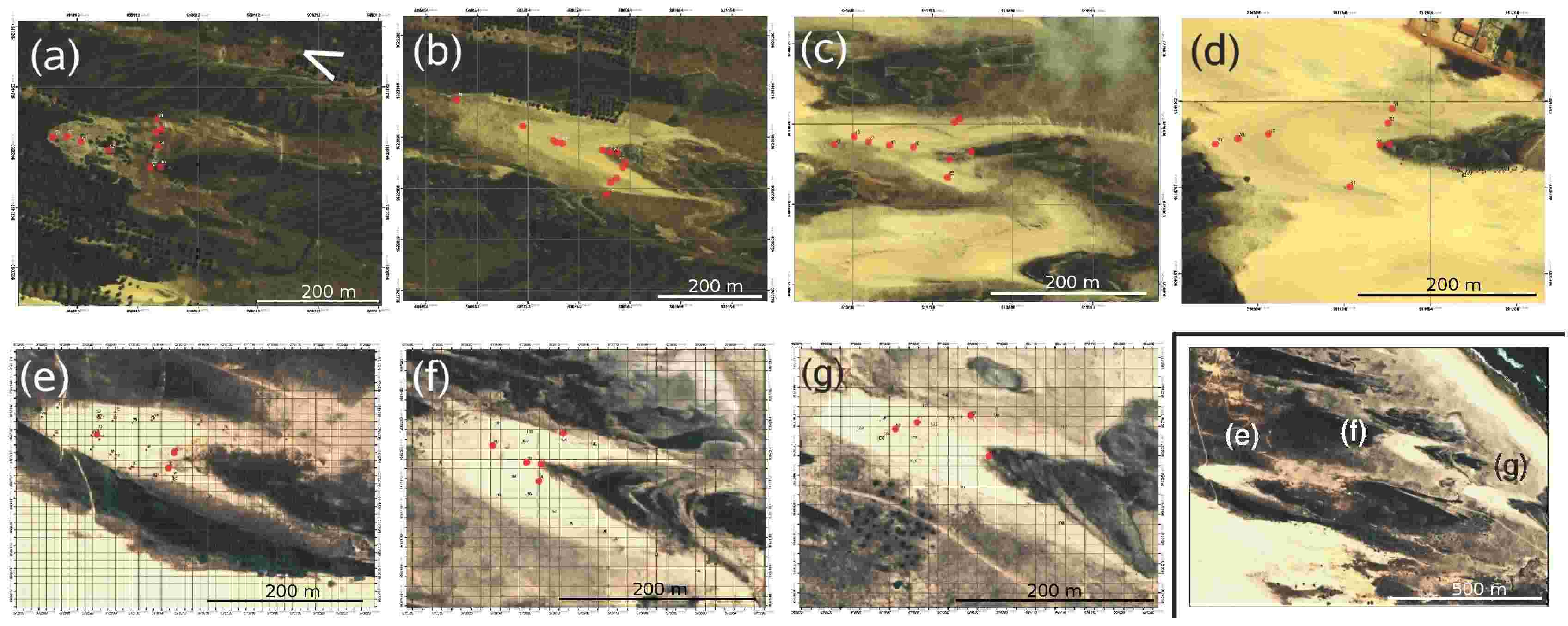}}
\caption{(Color online) {\bf Top:} Measured parabolic dunes from Paracuru (a) and (b), Pecem (c) and Taiba (d) in order of inactivation, the most inactive being (a) and the most active (d). {\bf Bottom:} the other three parabolic dunes from Iguape in order of inactivation, from the less active (e) to the most active (g). The bottom right shows a panoramic view of the cluster formed by these three dunes. The red (dark) dots on the dunes indicate the location where vegetation was measured (see Fig.~8). The North points up and wind blows from ESE as shown by the arrow in (a). The geographical coordinates of the dunes are: (a) $499912,9623553$, (b) $500754,9623000$, (c) $513790,9609650$, (d) $510984,9611800$, (e) $572930, 9567090$, (f) $573650,9567200$ and (g) $573990,9567000$.}
\end{figure}

\section{Materials and methods}

Since plants locally slow down the wind they can inhibit sand erosion as well as enhance sand accretion, as it is shown in Fig.~5. This dynamical effect exerted by plants on the wind, which is characterized by the drag force acting on them, is mainly determined by the frontal area density $\lambda\equiv A_f/A$, where $A_f$ is the total plant frontal area, i.e. the area facing the wind, of vegetation placed over a given sampling area $A$. Furthermore, the vegetation cover over a dune is defined by the basal area density $\rho_v\equiv A_b/A$, where $A_b$ is the total plant basal area, i.e. the area covering the soil, on $A$ \citep{Raupach92,Raupach93}. 

\begin{figure}[htb!]
\centering
\fboxrule=0.5mm
\framebox{\includegraphics[width=0.5\textwidth]{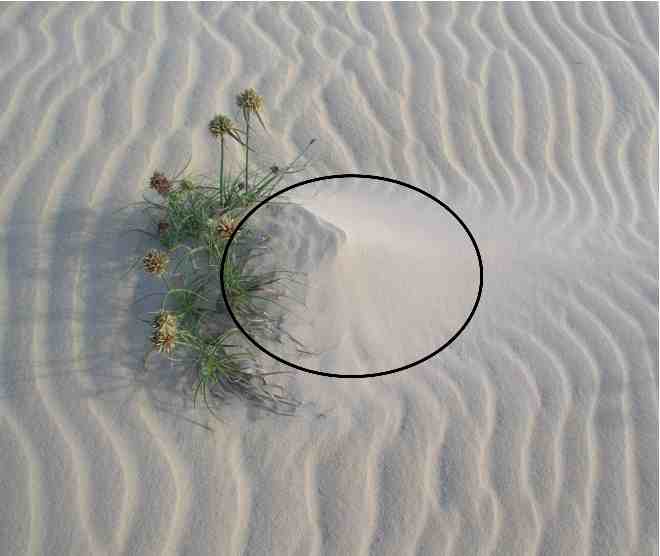}}
\caption{(Color online) Strong sand accumulation downwind a plant due to the local wind slow down. The wind velocity in the plant wake is below the threshold for sand transport, which can be deduced from the absence of ripples in that region. The effective shelter area is enclosed by the circle.}
\end{figure}

The distinction between both densities is crucial for the modeling of the vegetation effect over wind strength and thus sand transport \citep{Raupach92,Raupach93}. From Fig.~5 it is clear that plants act as obstacles that absorb part of the momentum transferred to the soil by the wind. As a result, the total surface shear stress $\tau$ can be divided into two components, a shear stress $\tau_v$ acting on the vegetation and a shear stress $\tau_s$ acting on the sand grains. When plants are randomly distributed and the effective shelter area for one plant (see Fig.~5) is proportional to its frontal area, the absorbed shear stress $\tau_v$ is proportional to the vegetation frontal area density $\lambda$ times the undisturbed shear $\tau_s$ \citep{Raupach92}. Using this it has been proposed that the fraction of the total stress acting on sand grains is given by

\begin{equation}
\label{taus}
\tau_s = \frac{\tau}{(1-m\rho_v)(1+m\beta\lambda)}
\end{equation}
where $\beta$ is the ratio of plant to surface drag coefficients and the constant $m$ is a model parameter that accounts for the non-uniformity of the surface shear stress \citep{Raupach93,Wyatt97}. The term $(1-m\rho_v)$ arises from the relation between the sandy and the total area.

Although $\rho_v$ and $\lambda$ can be calculated from direct measurements of the plants, in order to estimate the parameters $\beta$ and $m$ in Eq.~\ref{taus} we need a far more complex procedure since in this case we have to measure the drag forces acting on the plants and the shear stresses on the sand surface with and without plants. We visited the field location during the rain season when winds are exceptionally weak. The vegetation cover over any of the measured dunes includes at least six different species. We focused on measuring the static quantities like densities, rather than the dynamic ones like $\beta$ and $m$. In the section regarding the numerical solution of the model we use reported values from the literature for these two parameters \citep{Wyatt97}.

The basal and frontal area density, $\rho_v$ and $\lambda$, can be indirectly estimated from the local number $n_i$, basal area $a_{bi}$ and frontal area $a_{fi}$ of each species $i$ of plants over a characteristic area $A$ on the dune

\begin{eqnarray}
\label{rho}
\rho_v & \equiv & \frac{A_b}{A} = \frac{1}{A}\sum_i n_i a_{bi} \\
\lambda & \equiv & \frac{A_f}{A} = \frac{1}{A}\sum_i n_i a_{fi}
\label{lambda}
\end{eqnarray}
Figure 6 shows an sketch illustrating the local basal $a_b$ and frontal area $a_f$ of a given plant.

\begin{figure}[htb!]
\centering
\fboxrule=0.5mm
\framebox{\includegraphics[width=0.5\textwidth]{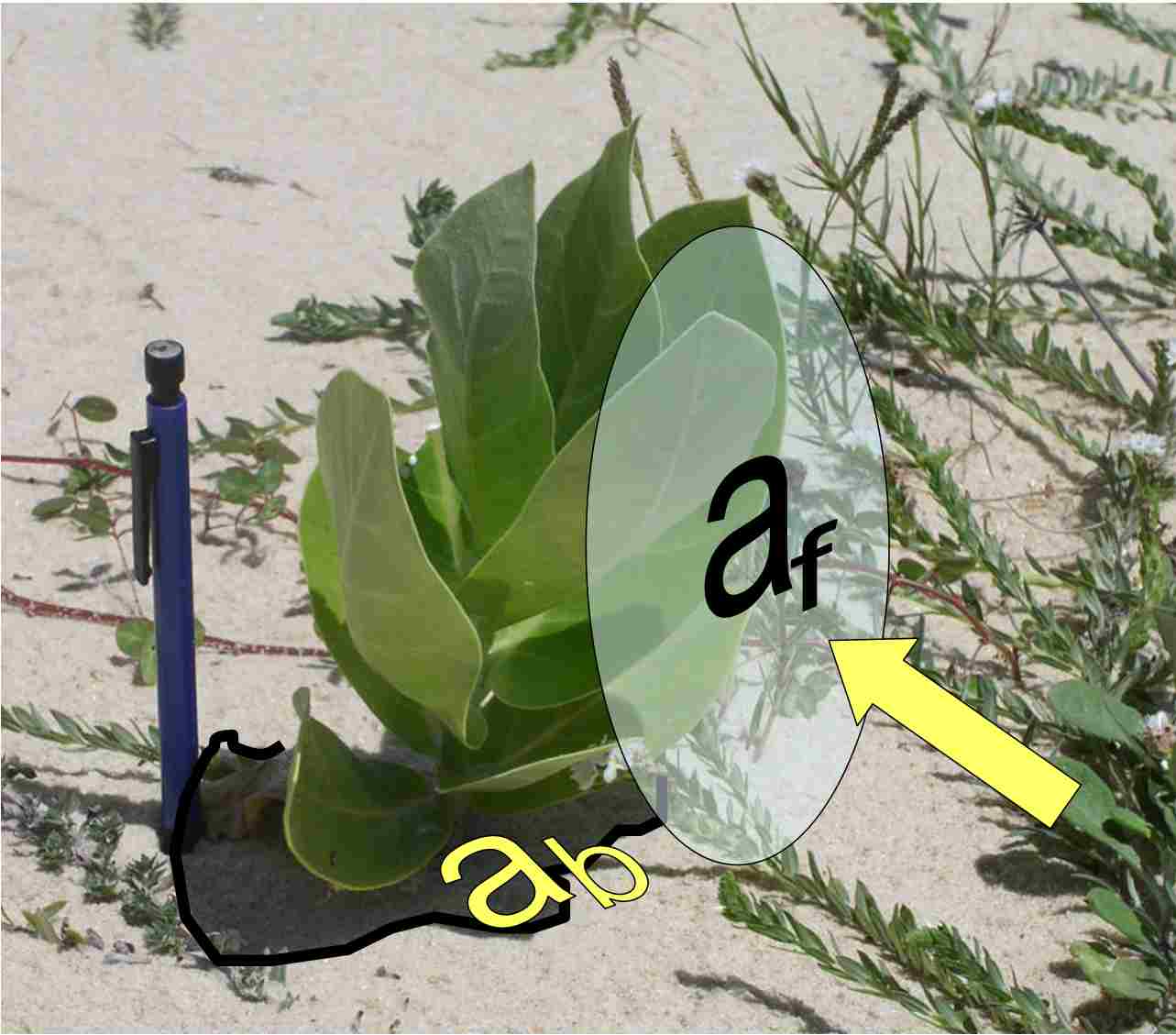}}
\caption{(Color online) Sketch of the local plant basal area $a_b$ covering the soil and the frontal area $a_f$ facing the wind.}
\end{figure}

\begin{figure}[htb!]
\centering
\includegraphics[width=0.53\textwidth]{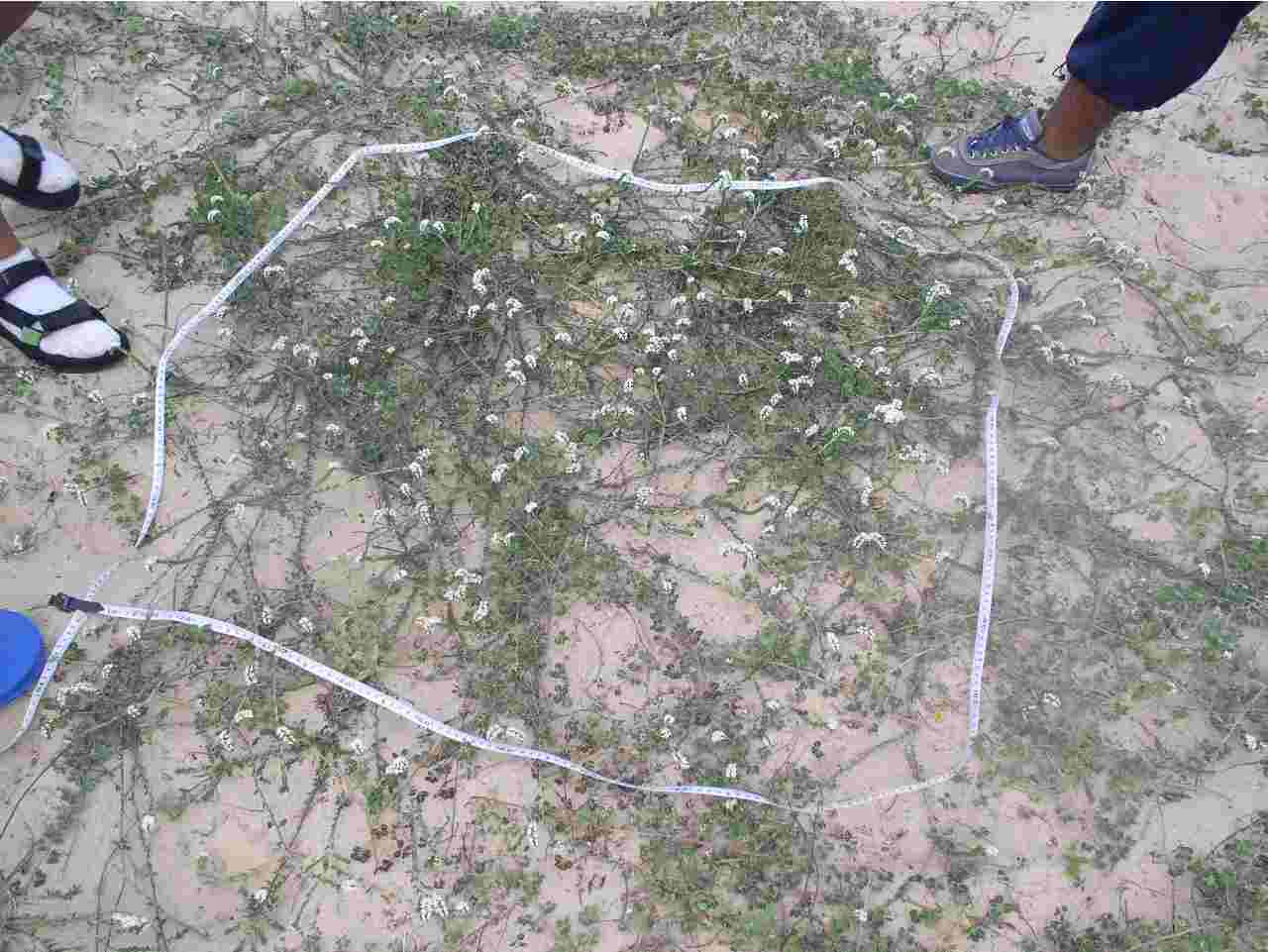}
\hspace{0.2cm}
\includegraphics[width=0.4\textwidth]{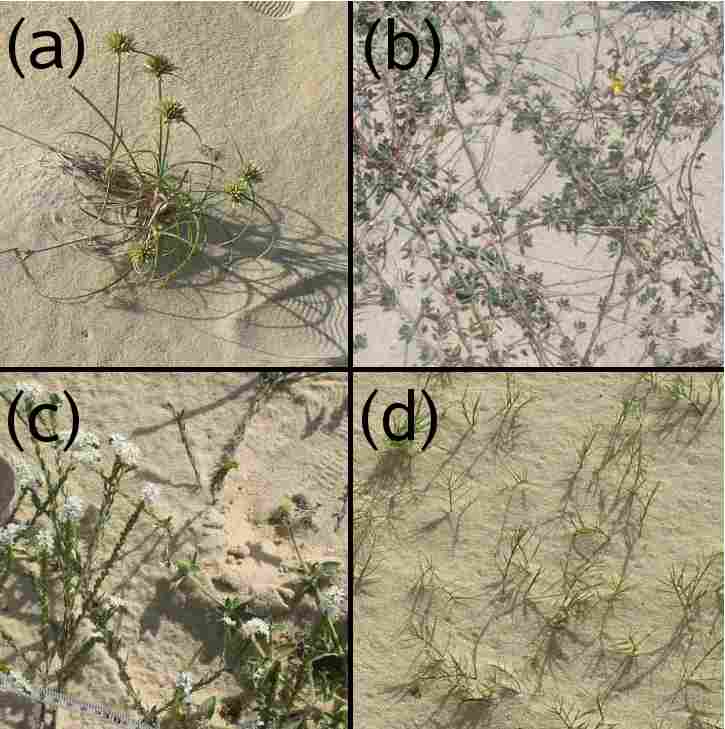}
\caption{(Color online) {\bf Left:} example of a sampling area used to measure the vegetation cover. {\bf Right:} four typical species of plants found in the region where measurements were performed: cyperus maritimus (a), chamaecrista hispidula (b), heliotropium polyphyllum (c) and sporobolus virginicus (d).}
\end{figure}

In order to measure the basal and frontal vegetation area density over the dunes shown in Fig.~4, which are mainly covered by grass, we select five to ten points along the longitudinal and transversal main axes of the parabolic dune (red dots in Fig.~4). On every point we identify each plant $i$ the number of times $n_i$ it appears in a study area $A=1$ m$^2$ and measure their characteristic length, height, number of leafs and leaf area (Fig.~7, left). Some species of the typical vegetation we found are shown on the right side of Fig.~7. Using the morphological information of each species $i$ we measure the fraction of the total leaf area of each plant that covers the soil $a_{bi}$, and the fraction that faces the wind $a_{fi}$.

In general, we found the same qualitative distribution of plants on all measured parabolic dunes (Fig.~8). The area between the arms of the dune is totally covered by plants, while their density reduces on the windward side where sand erosion is very strong and increases once again on the lee side, where most of the sand deposition occurs.

\begin{figure}[htb]
\centering
\includegraphics[width=1.0\textwidth]{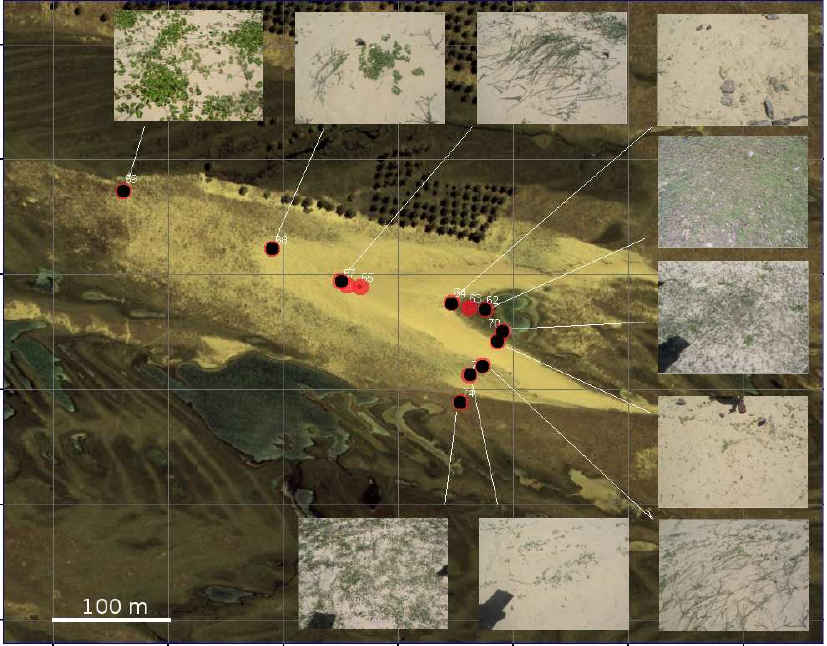}
\caption{(Color online) Satellite image of a typical measured parabolic dune with a close up of the vegetation growing in different places over the dune. Red (dark) dots indicate the places where vegetation data was collected. The North points up and wind blows from ESE.}
\end{figure}

\section{Results}

\subsection{Empirical vegetation cover on parabolic dunes}

By using the collected vegetation data we were able to calculate, through Eqs.~(\ref{rho}) and (\ref{lambda}), the basal and frontal plant density at some particular points on the dunes. The first interesting result is that both densities are proportional to each other (Fig.~9) with a proportionality constant $\sigma\approx 1.5$ with a reasonable dispersion in spite of the different dunes and types of vegetation. Therefore, the plant basal density, also called cover density, $\rho_v$, can be used to characterize the interaction between vegetation and the wind strength given by Eq.~\ref{taus}. This result agrees with measurements on Creosote bush reported in Wyatt et al. \citep{Wyatt97}. They also found the same value of $\sigma$ besides the enormous differences in the vegetation type, from bush in their case to grass in ours.

\begin{figure}[htb]
\centering
\includegraphics[width=0.7\textwidth]{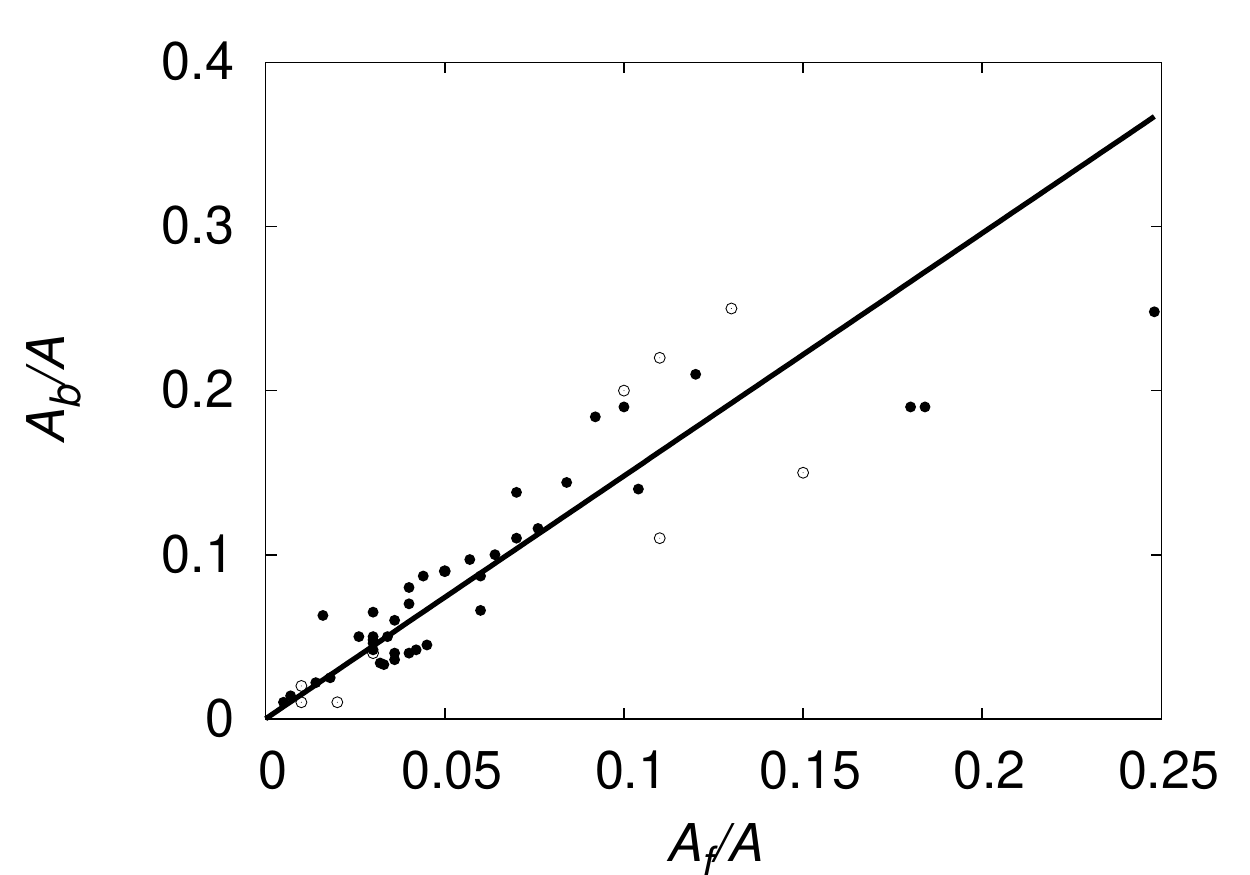}
\caption{Proportionality between the basal density $\rho_v\equiv A_b/A$ and the frontal area density $\lambda\equiv A_f/A$. The proportionality constant of the fit (solid line) is $A_b/A_f \equiv \sigma = 1.48$. Each point represents a different sampling area on the dunes situated in Iguape, shown in Fig.~4 bottom, ($\circ$) and the others dunes, Fig.~4 top, ($\bullet$).}
\end{figure}

In order to estimate the inactivation degree of the whole parabolic dune, based only on the plant cover density, we have to extend the few measured values of $\rho_v$ to the full dune body. The gray-scale of the satellite image (with a resolution 0.6 m/pixel) suggests that vegetation density determines the image darkness. Therefore, a crude approximation for the cover density at the dune is obtained by relating the density cover $\rho_v$ to its normalized image gray-scale value $C$. This value is defined as

\begin{equation}
C \equiv \frac{c-C_{min}}{C_{max}-C_{min}}
\end{equation}
where $c$ is the gray-scale value of a given point and $C_{min}$ and $C_{mac}$ are defined by the normalization conditions $\rho_v(C_{min})=1$ and $\rho_v(C_{max})=0$ respectively. These normalization conditions are obtained from those points in the image that we know are either bare sand or fully covered with plants.

\begin{figure}[htb]
\centering
\includegraphics[width=1.0\textwidth]{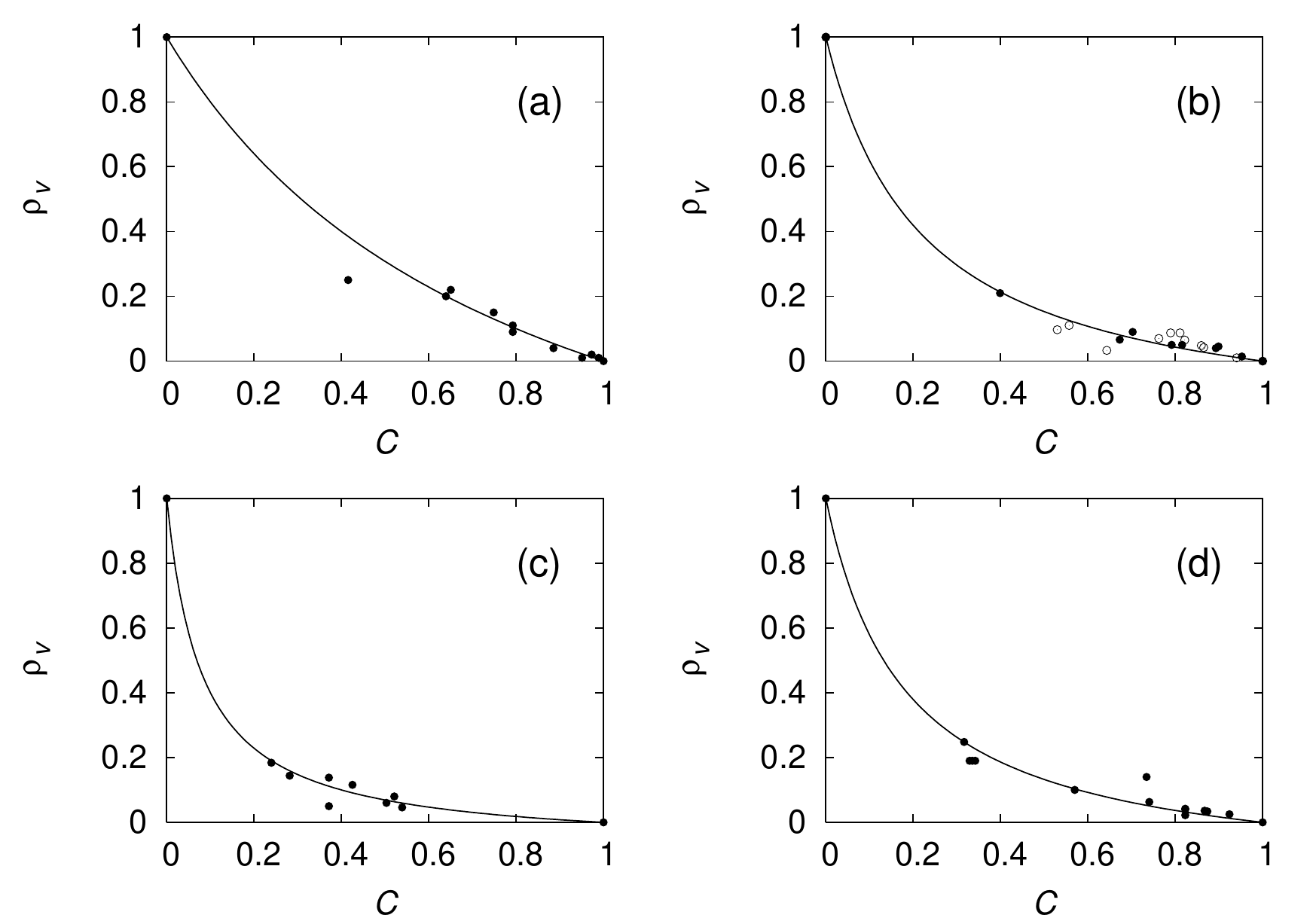}
\caption{Relation between the vegetation cover density $\rho_v$ and the normalized gray-scale value $C$ from the satellite image. Each point represents another sampling area and the plots correspond to: (a) the three dunes from Iguape (Fig.~4,e,f and g), (b) the two dunes from Paracuru, ($\bullet$) (Fig.~4a) and ($\circ$) (Fig.~4b), (c) the dune from Pecem (Fig.~4d) and (d) the dune from Taiba (Fig.~4c). The fit parameter $a$ in Eq.~\ref{C} has the values 0.8, 0.22, 0.08 and 0.18, respectively.}
\end{figure}

Figure 10 shows a clear correlation between both $\rho_v$ and $C$ for each image. By assuming that the cover density decreases linearly with $C$ when $C\rightarrow 1$ and that $\rho_v(C)$ is symmetric with respect to the main diagonal $\rho_v = C$, we propose the fitting curve

\begin{equation}
\label{C}
\rho_v(C) = a \frac{1-C}{a+C}
\end{equation}
where the fitting parameter $a$ changes for different images due to the alteration of the respective gray-scale.

With equation (\ref{C}) we can estimate the density cover over the whole parabolic dune. Figure 11 shows the resulting density cover calculated from the gray-scale of the images in Fig.~4. Yellow (light) represents free sand, while dark green (dark) represents total cover, and thus, total inactivation. With the help of the color-scale one identifies the zones where sand transport occurs. The windward side in the interior part is the most active part of the dune, as consequence of the erosion that prevents plants to grow. On the contrary, plants apparently can resist sand deposition since they accumulate clearly at the lee side and the crest of the dune. 

\begin{figure}[htb!]
\centering
\fboxrule=0.5mm
   \framebox{\includegraphics[width=1.0\textwidth]{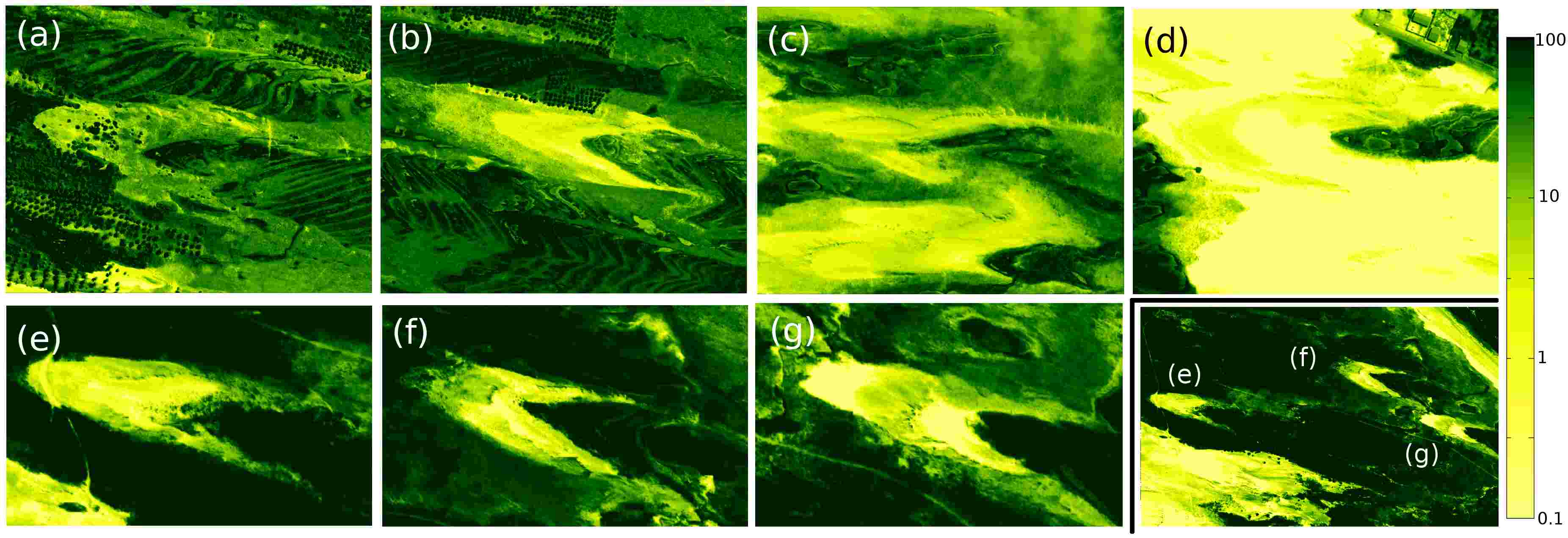}}
\caption{(Color online) Vegetation cover density on the seven measured parabolic dunes of the coastal zone of Brazil, depicted in Fig.~4. Yellow (light) represents no cover $\rho_v=0$ ($C=1$) and dark green (dark) total cover $\rho_v=1$ ($C=0$). The logarithmic color scale is in percentage.}
\end{figure}

Another important conclusion is that the degree of activation of a dune apparently depends on its distance from the place where it was born. The three dunes in Iguape (Fig.~11) are deactivated according to their distance from the sea shore, the most active being the one nearest to the sea (Fig.~11, bottom right). Similarly, the same occurs with other studied parabolic dunes.

\subsection{Vegetation cover on simulated parabolic dunes}

The numerical study of parabolic dunes has recently experienced new developments based on cellular automaton models \citep{Nishimori01,Baas02,Baas07}. Furthermore, we have recently proposed a continuum approach for the study of the competition process between vegetation growth and sand erosion \citep{Duran06b}. As was stated in the previous section, plants can locally slow down the wind, reducing erosion and enhancing sand accretion. On the other hand, sand is eroded by strong winds denuding the roots of the plants and increasing the evaporation from deep layers \citep{Tsoar02,Tsoar05}. As a result, there is a coupling between the evolution of the sand surface and the vegetation that grows over it, controlled by the competition between the reduction of sand transport rate due to plants and their capacity to survive sand erosion and accretion \citep{Tsoar05,Duran06b} (The full model is presented in the appendix).

\begin{figure}[htb!]
\centering
\fboxrule=0.5mm
   \framebox{\includegraphics[width=1.0\textwidth]{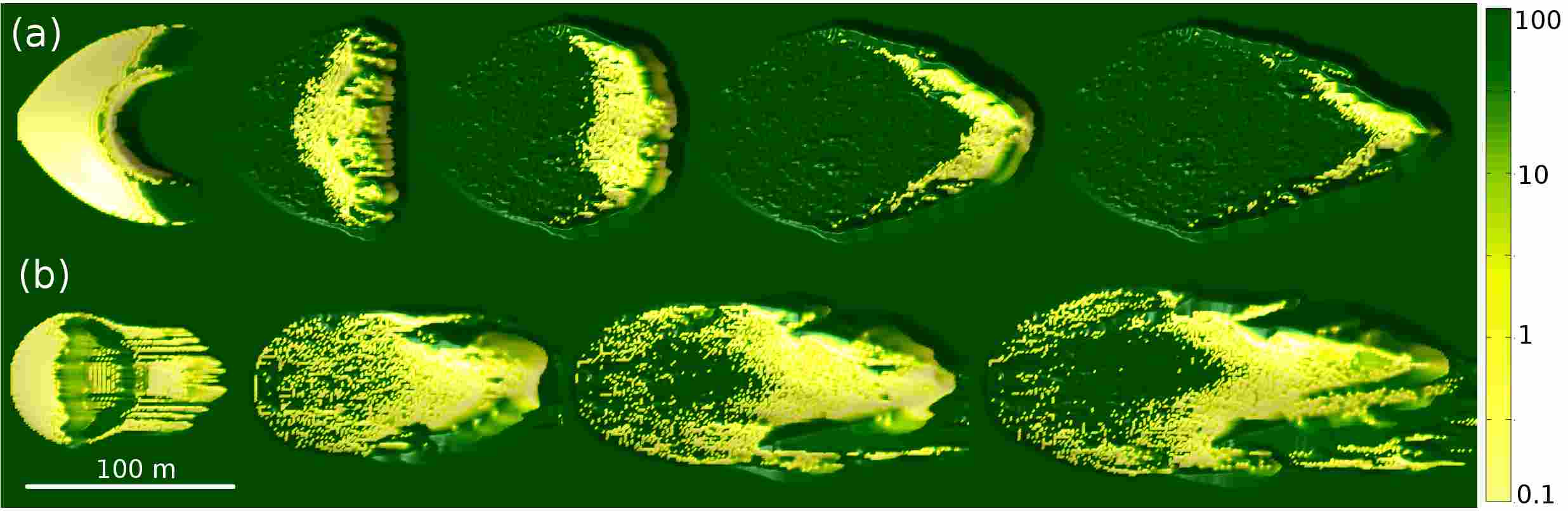}}
\caption{(Color online) {\bf (a)} Snapshots of the transformation of a barchan dune into a parabolic one. {\bf (b)} Evolution of a parabolic dune from a blow-out. The color-scale, in percentage, represents density cover, and, as in Fig.~11, yellow (light) represents no cover $\rho_v=0$ ($C=1$) and dark total cover $\rho_v=1$ ($C=0$). Wind blows from the left.}
\end{figure}

In order to compare with measurements we perform numerical calculations of parabolic dunes emerging from two different initial conditions. Figure 12a, shows the parabolic dune resulting from the evolution of a barchan dune under active vegetation growth \citep{Duran06b}, while in Fig.~12b the parabolic dune emerges from the evolution of a blow-out, i.e. a spot of bare sand within a vegetated surface, which are common in coastal systems \citep{Pye82,Hesp96}. 

\section{Discussion}

From Fig.~12, it is clear that for both initial conditions a parabolic dune evolves with the characteristic vegetated arms pointing upwind and a sandy windward side. Moreover, the convex and heavily eroded windward side finishes in a cut-edge vegetated crest, in sharp contrast with the barchan dune where the windward side is concave and the crest smoothly rounded. 

As can be seen in Fig.~12, it is in the morphology, and not the vegetation distribution, where the two parabolic dunes evolving from different initial conditions differ most. Although the dunes are of similar sizes, the windward side of the blow-out parabolic dune is more than twice the size of the windward side of the barchan-born parabolic dune. Furthermore, the blow-out dune is more elongated than the corresponding barchan-born one and concentrates a higher sand volume in its `nose'. This can be understood from their respective evolutions: on one hand, the former dune is growing from a spot of bare sand in the vegetated surface and its volume is increasing due to the sediment that is continuously added on the vegetated surface \citep{Hesp96,Pye82}. On the other hand, the barchan-born parabolic dune has a total sand volume fixed from the beginning and since this volume is distributed over a growing surface, the parabolic dune is continuously shrinking, particularly its `nose' \citep{Duran06b}.

After comparing the simulated parabolic dunes (Fig.~12) with the real ones from Brazil (Fig.~11) the similarity between the dune emerging from a blow-out and the measured ones seems evident. Figure 13 depicts one dune from Iguape and a simulated blow-out in their color-scale (Fig.~13a and c) and in the gray-scale corresponding to the quantitative vegetation cover density $\rho_v$ (Fig.~13b and d). Although the real dune has twice the size of the measured one, their morphology is very similar. Examples of their similarities are: their relative length and width; the position of the slip-face, that is sharper along the windward side than in the `nose', and the gentile slope in the windward side and the front of the `nose' compared to the step lee side. 

Regarding the distribution and density of the vegetation, both dunes also share strong similarities (Fig.~13b and d). The distribution of the vegetation is clearly divided into two regions, the windward side almost devoid of plants and the vegetated lee side (Fig.~14). As was discussed in the previous sections, this is a direct consequence of the competition between sand transport and vegetation growth, while on the heavily eroded windward side plant roots are uncovered and dried, on the lee side they survive sand accretion (see Fig.~14). 

\begin{figure}[htb!]
\centering
\fboxrule=0.5mm
   \framebox{\includegraphics[width=1.0\textwidth]{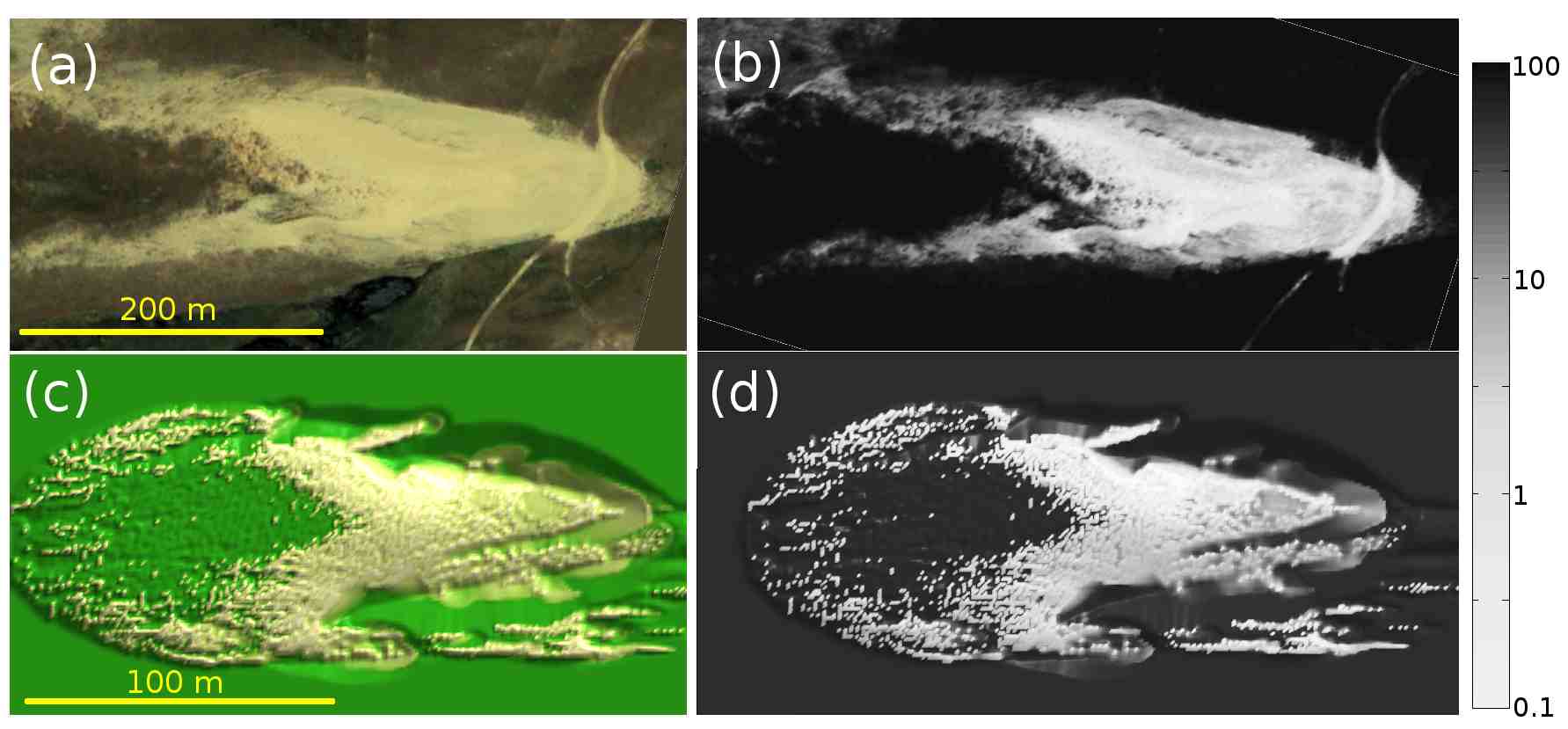}}
\caption{(Color online) Comparison between a real and a numerical parabolic dune. The morphology of the numerical dune emerging from a blow-out (c) is quite similar to the real one found in Iguape (a) (Fig.~4e). The vegetation cover over the numerical (d) and the real dune (b) also shows the same qualitative distribution.}
\end{figure}

\begin{figure}[htb!]
\centering
\fboxrule=0.5mm
   \framebox{\includegraphics[width=0.7\textwidth]{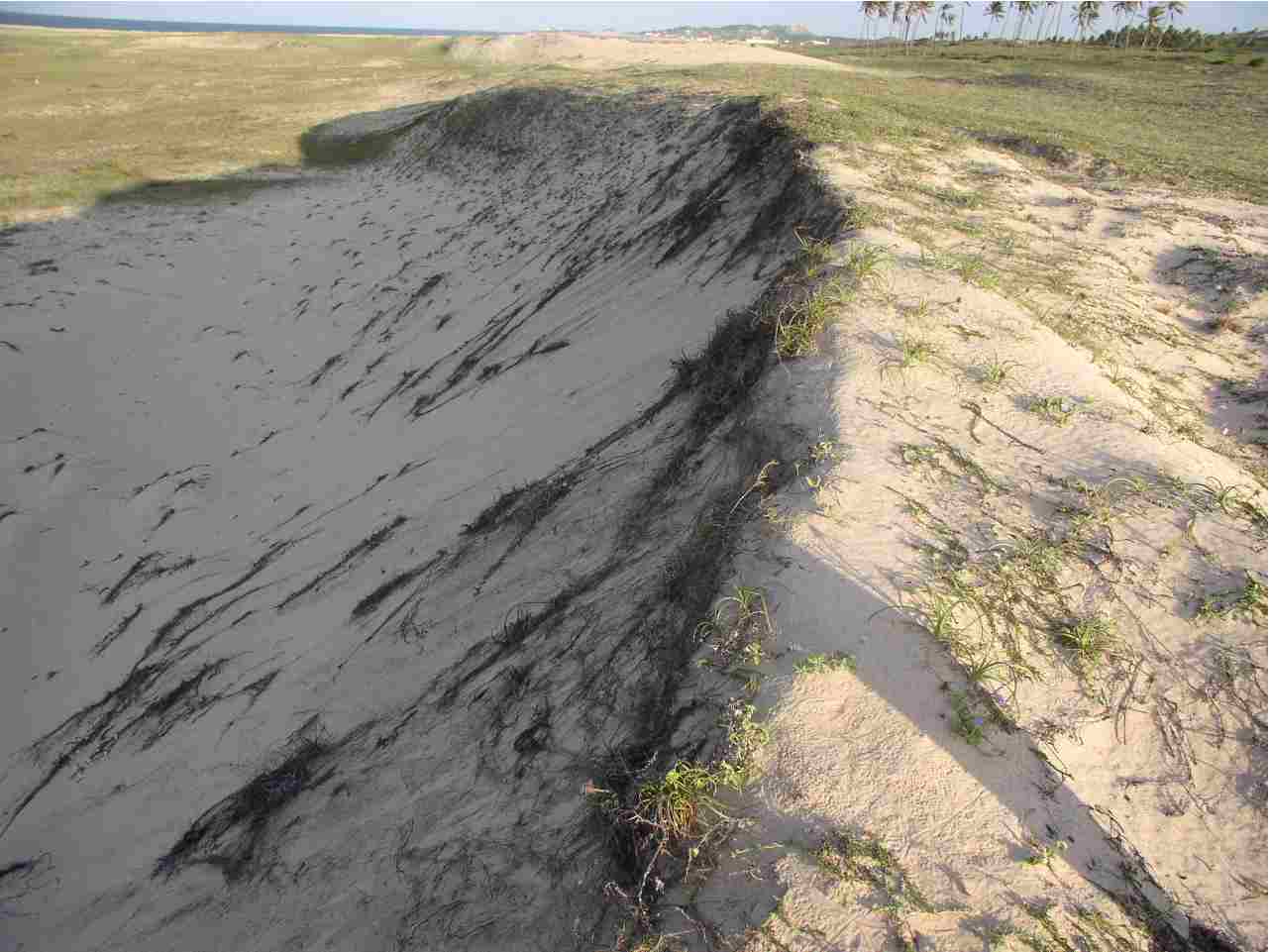}}
\caption{(Color online) The picture of one arm of a parabolic dune taken from the top of its `nose', illustrates the difference between the windward face, at the left, and the lee one, at the right. In the former the erosion has killed the vegetation, which remnants can be seen in dark, while in the lee side the plants are alive.}
\end{figure}

However, although the distribution of vegetation is qualitatively similar on both dunes, there is an important quantitative difference regarding the vegetation cover density: on the lee side of the simulated dune (Fig.~13d) the vegetation density reaches a far higher value, in fact very close to one, than in the real dune (Fig.~13b). This could be consequence of the way the effect of the vegetation on the sand transport is modeled. We assume that the shear stress partition, i.e. the wind slow down, is the only way plants can affect the sand transport, but there are other effects. For instance, plants act as physical barriers for the saltating grains that are carried by the wind. Plants can trap them by direct collisions as well as by reducing the wind stress. In this case, a given vegetation density can be more effective in avoiding soil erosion than when one only considers the wind slow down. This can explain why the lee side on both dunes can have the same protecting role while having different vegetation density cover.

\section{Conclusions}

We presented measurements on real parabolic dunes along the coast of Brazil, concerning their shape and the vegetation cover on them. The vegetation cover over a dune was estimated by the number and size of plants in a characteristic area of the dune. To do so we identify the species of plants present on the dune and count the number of times they appear in the study area and measure their characteristic length, height and total leaf area. 

By using the vegetation data we were able to calculate the plant cover
density at particular points on the dunes and compare it with the gray scale of the satellite image. Doing so we found a relation between both the vegetation cover density and the image gray-scale which leads us to estimate the density cover on the whole parabolic dune. Then we compare the empirical vegetation cover data with the result of our simulations to validate quantitatively the distribution of vegetation over the dune. We conclude that the model indeed captures the essential aspects of the interaction between the different geomorphological agents, i.e. the wind, the surface and the vegetation. Furthermore, it gives arguments in favor of the possible origin of parabolic dunes on the Brazilian coast as coming from a blow-out, rather than from an early active coastal barchan dune system. Through this we could estimate the degree of inactivation of the dune and reconstruct the previous dune history.

\section*{Acknowledgment}

We wish to thank LABOMAR in Fortaleza for the kind help during the field work. This study was supported by the Volkswagenstiftung, the Max Plank Prize and the DFG.


\appendix

\section{Model for vegetated dunes}
The vegetated dune model consists of a system of continuum equations in two space dimensions that combines a description of the average turbulent wind shear force above the dune including the effect of vegetation, a continuum saltation model, which allows for saturation transients in the sand flux, and a continuum model for vegetation growth \citep{SauermannKroy01,KroySauermann02,SchwaemmleHerrmann05,Duran05,Duran06b}. The model can be sketched as follows: (i) first, the wind over the surface is calculated with the model of \citet{Weng91} that describes the perturbation of the shear stress due to a smooth hill or dune. The Fourier-transformed components of this perturbation are

\begin{eqnarray}
\tilde{\hat{\tau}}_x & = & \frac{\tilde h k^2_x}{|\vec k|}\frac{2}{U^2(l)}\left\{ -1 + \left( 2\ln \frac{l}{z_0} + \frac{|\vec k|^2}{k^2_x} \right)\sigma\frac{K_1(2\sigma)}{K_0(2\sigma)}\right\} \\
\tilde{\hat{\tau}}_y & = & \frac{\tilde h k_x k_y}{|\vec k|}\frac{2}{U^2(l)} 2\sqrt{2}\sigma K_1(2\sqrt{2}\sigma)
\end{eqnarray}
where x and y mean, respectively, parallel and perpendicular to the wind direction, $\sigma = \sqrt{i L k_x z_0/l}$, $K_0$ and $K_1$ are modified Bessel functions, and $k_x$ and $k_y$ are the components of the wave vector $\vec k$, i.e. the coordinates in Fourier space. $\tilde h$ is the Fourier transform of the height profile, $U$ is the vertical velocity profile which is suitably non-dimensionalized, $l$ is the depth of the inner layer of the flow, and $z0 = 1.0 mm$ is the aerodynamic roughness. $L$ is a typical length scale of the hill or dune and is given by $1/4$ the mean wavelength of the Fourier representation of the height profile. (ii) next, the effect of the vegetation over the surface wind -shear stress partitioning- is calculated by Eq.~\ref{taus}, which gives the fraction $\tau_s$ of total stress $\tau$ acting on the grains. (iii) the sand flux is calculated using the shear velocity $u_* = \sqrt{\tau_s/\rho_{\text{fluid}}}$, where $\rho_{\text{fluid}} = 1.225 kg/m^3$ is the air density, with the equation

\begin{equation}
\nabla\cdot\vec q = \frac{|\vec q|}{l_s}\left( 1 - \frac{|\vec q|}{q_s} \right) ,
\end{equation}
where $q_s = (2 v_s \alpha/g) u^2_{*t} [ (u_*/u_{*t})^2 - 1 ]$ is the saturated flux and $l_s = (2 v^2_s \alpha/\gamma g)/[ (u_*/u_{*t})^2 - 1 ]$
is called saturation length; $u_{*t} = 0.22 m/s$ is the minimal threshold shear velocity for saltation and $g = 9.81 m/s^2$ is gravity, while $\alpha = 0.43$ and $\gamma = 0.2$ are empirically determined model parameters and the mean grain velocity at saturation, $v_s$, is calculated numerically from the balance between the forces on the saltating grains; (iv) the change in surface height $h(x, y)$ is computed from mass conservation: $\partial h/\partial t = -\nabla\cdot\vec q/\rho_{\text{sand}}$, where $\rho_{\text{sand}} = 1650 kg/m^3$ is the bulk density of the sand; (v) if sand deposition leads to slopes that locally exceed the angle of repose, $34^o$, the unstable surface relaxes through avalanches in the direction of the steepest descent, and the separation streamlines are introduced at the dune lee. Each streamline is fitted by a third order polynomial connecting the brink with the ground at the reattachment point, and defining the ``separation bubble", in which the wind and the flux are set to zero. (vi) finally, the vegetation growth rate is calculated from the surface change using the phenomenological equation 

\begin{equation}
\frac{dh_v}{dt} = V_v\left(1 - \frac{h_v}{H_v}\right) - \left|\frac{\partial h}{\partial t}\right|
\end{equation}
where it is assumed that a plant of height $h_v$ can grow up to a maximum height $H_v$ with an initial rate $V_v$. To close the model, the basal area density introduced in Eq.~\ref{taus} is just $\rho_v = (h_v/H_v)^2$, and the frontal area density $\lambda = \rho_v/\sigma$ as Fig.~9 shows.
The model is evaluated by performing steps i) through vi) computationally in an iterative manner.


\end{document}